\documentclass[runningheads]{llncs}
\usepackage{graphicx}
\begin{document}
\title{Toward Best Practices for\\Explainable B2B Machine Learning }
%
%
\author{Kit Kuksenok\inst{1}}
\authorrunning{K. Kuksenok}
%
\institute{jobpal Ltd., Berlin, Germany \\
\email{kit@jobpal.ai}}
\maketitle              
\begin{abstract}
To design tools and data pipelines for explainable B2B machine learning (ML) systems, we need to recognize not only the immediate audience of such tools and data, but also (1) their organizational context and (2) secondary audiences. Our learnings are based on building custom ML-based chatbots for recruitment. We believe that in the  B2B context, ``explainable'' ML means not  only a system that can ``explain itself'' through tools and data pipelines, but  also enables its domain-expert users to explain it to other stakeholders.

\end{abstract}

\section{Introduction}
Based on experiences with implementing conversational agents in the recruitment domain using  machine learning (ML), we outline key considerations for best  practices for making enterprise (B2B) ML systems  explainable. Recruitment chatbots mediate communication between job-seekers and recruiters by exposing ML data to recruiter teams, and  using  it to provide answers to frequently-asked questions, and notify job-seekers when new jobs fitting their  search criteria are available. Although transparency (being  ``honest and transparent when explaining why something doesn't work'') is a core chatbot design recommendation \cite{dialogflowerrors}, the most commonly available higher-level platforms~\cite{canonico2018comparison} do not provide robust ways to understand error and communicate  its implications. Errors are especially difficult to understand, communicate, and resolve because they may span and combine UX, ML, and software issues.  \textit{Interpretability} is a  challenge beyond the chatbot  domain, and is a prerequisite for trust in both individual predictions and the overall model~\cite{ribeiro2016should}. There are also many organizational and institutional challenges to ``black-box'' systems, including  ML systems and systems using AI technologies~\cite{pasquale2015black}. We believe that B2B ML is ``explainable'' when:

\begin{itemize}
    \item there are internal tools and data pipelines that support domain-experts (e.g., project managers) to interpret the ML systems, including correctly assessing and addressing common errors
    \item domain-experts internally are enabled by  these resources to synthesize their understanding and  communicate it to external stakeholders (e.g., recruitment  teams on the side of  the  client)
\end{itemize}

\begin{figure}
    \includegraphics[width=210pt]{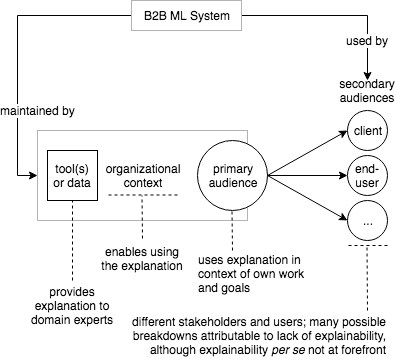}
    \caption{Whereas tools and data pipelines can be designed to provide an explanation to trained domain experts, the \textbf{organizational context} enables those domain experts to use the explanation. Furthermore, the aim of the explanations does not end at the internal, domain-expert primary audience of  these tools.}
    ~\label{fig:exai}
\end{figure}

\section{Our Approach}

To  achieve this aim through our internal tools and organizational culture, we recognize that (1)  the  organizational context  is vital to the usefulness of tools and data pipelines to domain-experts, and (2) the \textit{direct} users of tools and data pipelines to explain ML are not the only audience that it helps to deliberately design for (see  also Fig.~\ref{fig:exai}). In existing  work on enabling domain-experts to interact  with  machine learning\footnote{For example, in ~\cite{cavallo2019clustrophile} the  authors use  both visual and textual encodings  of data, and guide exploratory analysis using ideas that we  have drawn upon in internal tools.},  the focus is typically on supporting the informational tasks of the domain-experts  using  the  system, which are the  system's \textit{primary} audience. At jobpal, we have also developed tools and data pipelines for visual,  speculative, and complex data analysis, but in developing these we found it useful  to be explicitly  aware of the \textit{secondary} audience,  and of  the organizational context of internal tool use.

Consider the following examples of two common questions that drives the interaction between the primary and secondary audiences\,---\,downstream from internal resources that aim to explain the ML system.

\subsection{``Is it worth it?''}

Smart conversational agents are increasingly used across business domains ~\cite{jain2018evaluating}. We focus on recruitment chatbots  that connect recruiters and job-seekers. The recruiter teams we  work with are motivated to build and maintain chatbots  that provide answers to frequently  asked questions (FAQs) based on ML/NLP datasets for reasons of scale and accessibility. Our enterprise clients may have up to $100K$ employees, and commensurate hiring rate. We have found that almost $50\%$ of end-user (job-seeker) traffic occurs outside of working hours ~\cite{jobpal2019when}, which is consistent  with  the anecdotal reports of  our  clients that using the chatbot  helped reduce email and ticket inquiries of common FAQs.

In the early stages of a client project, or before a project begins,  the question arises: ``is it worth it?'' Partial automation of  even small tasks brings with it unexpected discoveries, which may or may not map onto organizational or team goals. At this  stage, the best  strategy is to discourage comparison of the ML-based chatbot  to  a person, as well as to remind potential clients that it is not AI, and it cannot  solve every  problem. Rather, it  is an experiment, and,  in the  best case, it gives more  concrete  ideas about how the recruitment team might change an aspect of their work. For example, it may reveal  that the anticipated behavior or  needs of the end-users do not match the  actual behavior or  needs. It may also reveal how much more consistent responding to  those  questions can be relative to  branding guidelines.

To frame it as an experiment that requires active  participation in design and implementation from all stakeholders benefits from a well-supported deployment and prototyping cycle, which is familiar to all internal  stakeholders.  The question ``is it worth it?'' cannot  be  answered  without experimentation in the specific context.

\subsection{``What's wrong with it?''}

To support the prototyping and deployment process, project managers undertake data quality management tasks that require using  internal tools developed for this task. In this example, the  primary audience  includes a project manager,  the secondary audience includes the  client to whom the  project manager explains a  new feature or a surprising  behavior. Whether the project manager finds the explainable ML tools or data pipelines usable  depends on the organizational culture: this includes providing support, incorporating feedback, and valuing the time they invest in engaging  with these internal resources.

Besides the usefulness of these resources to help the project  manager build \textit{their own} understanding, the hope is that these tools enable them to \textit{synthetize} their understanding with client  (secondary audience) needs, and \textit{communicate} in a way that effectively  builds understanding of machine learning into an already complex interaction.  Recognizing the additional work and complexity of applying  understandable ML with respect to various additional stakeholder groups can help to build more robust  tools and surrounding organizational culture that  enables  it.

\section{Implications for  Design}

 We suggest the following question to help guide development of  explainable  ML systems:

\begin{itemize}
    \item Who are the primary and secondary audiences?
    \item What are the technological resources available to  support internal stakeholders in building their understanding of relevant ML systems?
    \item What secondary  users' goals and needs that the  primary users will support? 
    \item In what ways is the  work of understanding and  explaining ML recognized and supported  organizationally?
\end{itemize}

It is crucial to think, at every step of designing, developing, and maintaining an ML system, about how it can be made explainable. This position paper asserts that in the  B2B context, ``explainable'' ML means not  only a system that can ``explain itself'' through tools and data pipelines, but  also enables its domain-expert users to explain it to other stakeholders.

\section{Acknowledgements}

This position paper draws upon conversations with Klaudia Niedba\l{}owska and Nina Pra{\ss}, project managers  who use internal tools for communicating data  quality issues and recommendations to clients with ongoing projects; and Chris Raw, product strategist and talent acquisition expert who provided  insights on communicating the capabilities of recruitment chatbots to potential clients or in early stages of a project.

\end{document}